\documentclass{vgtc}                          %

\graphicspath{{figures/}{pictures/}{images/}{./}} %

\usepackage{times}                     %

\usepackage{tabu}                      %
\usepackage{booktabs}                  %
\usepackage{lipsum}                    %
\usepackage{mwe}                       %

\usepackage{mathptmx}                  %

\usepackage{enumitem}

\usepackage{amsmath}                  %

\onlineid{0}

\vgtccategory{Research}

\vgtcinsertpkg

\usepackage[most]{tcolorbox}
\definecolor{mybg}{HTML}{F2F9FD}
\definecolor{myframe}{HTML}{7DA0B0}
\definecolor{NavyBlue}{RGB}{0,0,128}

\title{Visual Analytics for Causal Reasoning from Real-World Health Data}

\author{Arran Zeyu Wang\thanks{e-mail: zeyuwang@cs.unc.edu}\\ %
        \scriptsize UNC Chapel Hill %
\and David Borland\thanks{e-mail: borland@renci.org}\\ %
     \scriptsize RENCI, UNC Chapel Hill %
\and David Gotz\thanks{e-mail: gotz@unc.edu}\\ %
     \scriptsize UNC Chapel Hill}

\abstract{
    The increasing capture and analysis of large-scale longitudinal health data offer opportunities to improve healthcare and advance medical understanding. However, a critical gap exists between (a)---the observation of patterns and correlations, versus (b)---the understanding of true causal mechanisms that drive outcomes. An accurate understanding of the underlying mechanisms that cause various changes in medical status is crucial for decision-makers across various healthcare domains and roles, yet inferring causality from real-world observational data is difficult for both methodological and practical challenges. This Grand Challenge advocates increased Visual Analytics (VA) research on this topic to empower people with the tool for sound causal reasoning from health data. We note this is complicated by the complex nature of medical data---the volume, variety, sparsity, and temporality of health data streams make the use of causal inference algorithms difficult. Combined with challenges imposed by the realities of health-focused settings, including time constraints and traditional medical work practices, existing causal reasoning approaches are valuable but insufficient. We argue that advances in research can lead to new VA tools that augment human expertise with intuitive and robust causal inference capabilities, which can help realize a new paradigm of data-driven, causality-aware healthcare practices that improve human health outcomes.
} %


\begin{document}

\firstsection{Introduction}

\maketitle

\label{sec-intro}

The increasingly embraced medical data revolution provides many promising opportunities to improve health outcomes through data-driven practices~\cite{dash2019big}, but also introduces significant new challenges. Visual analytics has been seen as a powerful tool to address many of these challenges~\cite{gotz2016data}, and much progress has been made. However, one crucial aspect remains underexplored (for both health and the visual analysis discipline in general~\cite{borland2024using,hullman2021designing}): \textit{visual analytic tools for causal reasoning}.

\begin{figure}[htbp]
    \centering
    \vspace{-1em}
    \includegraphics[width=0.4\linewidth]{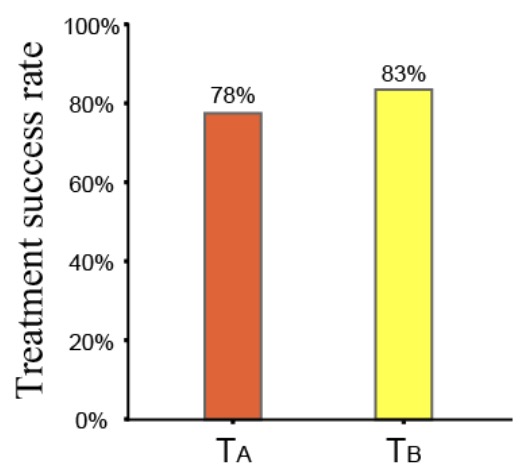}
    \vspace{-0.75em}
    \caption{A traditional bar chart visualization of treatment A ($T_A$) and treatment B ($T_{B}$) for kidney stone treatment data~\cite{charig_comparison_1986}.}
    \label{fig:kidney}
   \vspace{-1em}
\end{figure}

Causal understanding and inference are central to many medical decisions. Doctors choose interventions with the aim of causing improvements to a patient's condition. Patients choose treatments with the goal of causing benefits to their health status. Public health professionals choose policies with the aim of causing improvements to the health outcomes for a given population. 
Yet many visualizations of health data rely on representations of patterns and correlations that leave any causal inferences to the user without proper visual support.

For example, consider a bar chart showing data for two kidney stone treatments, $T_A$ and $T_B$, where $T_B$ exhibits a higher success rate (\autoref{fig:kidney}). A decision maker looking at this chart might understandably come to the conclusion that the data support the use of $T_B$.
However, omitted from the design of this chart is a critical, high-stakes question: Why? Was it the treatment itself that caused an improvement for $T_B$ patients, or perhaps $T_B$ patients were different from those that received $T_A$ in a meaningful way (younger, healthier, milder disease state)? In this case, we see that $T_B$ is correlated with better outcomes, but is it the cause? \textbf{Incorrectly mistaking correlation for causation} can lead to misguided decisions that worsen rather than improve health outcomes.

In the previous example, a closer examination of the kidney stone treatment data (from a real-world case~\cite{julious1994confounding}) shows that the populations receiving the two treatments $T_A$ and $T_B$ were in fact quite different in how their condition presented (small vs. large stone size). In a classic case of Simpson's Paradox, the data show that $T_B$ had a higher efficacy rate overall while $T_A$ performed better for both small and large stone populations. This simple but compelling example provides a warning regarding the potential of confounding variables to mislead us as we seek data-based evidence for our decisions.

\begin{table}[htbp]
    \centering
    \vspace{-0.5em}
    \caption{Success rates 
    of kidney stone treatments ($T_A$ and $T_B$) for small and large stones~\cite{charig_comparison_1986}. Values in \textbf{bold} imply that $T_B$ is more effective overall while  $T_A$ is more effective for both small and large stones.}
    \resizebox{0.3\textwidth}{!}{
    \begin{tabular}{|c|c|c|}
    \hline
    Stone Size & $T_A$ & $T_B$   \\
    \hline
    Small & \textbf{93\%} (81/87) & 87\% (234/270)  \\
    Large & \textbf{73\%} (192/263) &	69\% (55/80)  \\
    \hline
    All & 78\% (273/350) & \textbf{83\%} (289/350)  \\
    \hline
    \end{tabular}
    }
   \vspace{-0.5em}
    \label{tab:treatment}
\end{table}

Meanwhile, medical data can be enormously complex, with potentially hundreds of thousands of non-independent variables to consider (the ICD-10 system alone has approximately 70,000 distinct codes),
Disentangling the causal relationship between any single event (e.g., a particular treatment decision) and a desired outcome becomes enormously challenging. This becomes even more difficult with the introduction of advanced statistical and AI models which can make the connection between specific factors and outcomes even more opaque.

Visual Analytics (VA) offers a powerful way forward, not just for data exploration, but as a crucial bridge for \textbf{insight-driven causal inference and validation}.
VA is more than just data visualization; if designed to support causal reasoning, VA could serve as a bridge between users and the underlying data and algorithms. It could enable a collaborative process where users (e.g., clinicians, patients, policymakers) could partner with intelligent systems via a ``causal conversation'' with the data.
This position paper calls upon the visual analytics community to tackle one of the most profound challenges in data-driven healthcare: \textbf{uncovering causal relationships from real-world data to improve health decision-making}.

\label{sec-challenge}

\section{The Frailty of Statistical Causal Models}

Existing visualization methods~\cite{wang2015visual, wang2017visual, guo2023causalvis, fan2024visual} heavily rely on the use of expert-defined directed acyclic causal graphs~\cite{pearl2009causality}.
However, this approach could often face challenges in real-world healthcare:

\noindent \textbf{Computation}:
Automatically discovering a causal graph from data is an \textit{NP-hard} problem~\cite{gimenez2008complexity}.
For a typical clinical dataset, this is not just a theoretical concern.
Consider a moderately complex dataset from an EHR system with thousands of dimensions and sometimes yields 1 million~\cite{gotz2016data} per patient—a mix of demographics, diagnoses, lab results, medications, and vital signs taken over time.
The number of potential relationships between these variables explodes into the trillions, making the exhaustive computation of a reliable causal graph practically impossible.

\noindent \textbf{Comprehension}:
Even if we could compute such a causal graph, what would a patient, a clinician, or a health analyst do with it?
A visualization of a causal graph with a number of nodes and a number of directional causal links could possibly be an incomprehensible ``\textit{hairball}'' general users, who are usually not experts in statistics.
It therefore may fail to provide the actionable insight that is the ultimate goal of the analysis~\cite{borland2024using}.
The challenge is to \textit{invent novel visual analytics paradigms that open up this black box}, allowing users to scrutinize, understand, and even steer the process of creating comparable cohorts, thus understanding causal graphs.

\begin{tcolorbox}[
    colback=mybg, 
    colframe=myframe, 
    title={\textbf{Future Direction}},
    fonttitle=\bfseries,
    left=1mm, right=1mm, top=1mm, bottom=1mm,
    boxsep=1mm
]
\begin{itemize}[left=0pt]
    \item \textbf{Transparent Visual Causal Modeling.} A key opportunity for VAHC is to create visual analytics tools that make the causal modeling process transparent and interpretable---allow users to interactively build causal models, visualize the impact of certain variables, and understand the trade-offs made by statistical methods.
\end{itemize}
\end{tcolorbox}

\section{Limitations beyond Causal Graphs}

Statistical techniques like propensity scoring~\cite{haukoos2015propensity} have proven useful for causal inference by creating comparable groups from observational data.
With this idea, counterfactual visualizations~\cite{wang2024empirical, wang2024framework, wang2024beyond} were proposed to help users reason causality by constructing ``what if'' data subsets.
They lower the barrier to entry by not requiring causal models and statistical expertise.
However, the processes and assumptions often remain opaque and challenging to clinical users.

\noindent \textbf{Temporality}:
Causality in healthcare is fundamentally tied to sequences and time---all use cases in health data can be relevant to time~\cite{borland2021enabling}.
While seemingly straightforward, effectively visualizing temporal data to support causal reasoning is exceptionally difficult.
In such conditions, a meaningful counterfactual question is rarely ``What if the patient had been given Treatment B?'' but rather, ``What if Treatment B had been administered after the fever spiked on day 3, instead of Treatment A?''
Initial effort~\cite{jin2020visual} has been made to deal with event sequence data; however, it heavily relied on even more complex causal models, and may implicitly rely on \textit{temporal precedence as a proxy for causation}.
For example, such a visualization might clearly show that patients receiving a new therapy were less likely to be readmitted to the hospital.
However, it may fail to visually account for a critical hidden confounder: perhaps the new therapy was only given to healthier patients in the first place.
The visualization, by merely ordering events, can unintentionally reinforce a fallacious causal conclusion.

\noindent \textbf{Dynamic real-world scales}:
While promising, many proposed techniques have only been demonstrated on simplified, low-dimensional datasets.
With a 1000+ dimension patient record, a simple user interface may become unusable, e.g., which of the thousand features should the user be prompted to change?
How can the system guide the user to plausible or meaningful alternative scenarios without overwhelming them with infinite choices?
Existing systems lack the guidance and navigational functionalities needed to \textit{make causal inference processes feasible and meaningful at this scale}.

\begin{tcolorbox}[
    colback=mybg, 
    colframe=myframe, 
    title={\textbf{Future Direction}},
    fonttitle=\bfseries,
    left=1mm, right=1mm, top=1mm, bottom=1mm,
    boxsep=1mm
]
\begin{itemize}[left=0pt]
    \item \textbf{Visualizing Causal Effects in Temporal Treatment Pathways.} 
    An urgent challenge for VAHC is to possibly connect visual causal inference with event sequence visualization, moving from a logic of ``\emph{this followed that}'' to ``\emph{this happened instead of that, considering these other factors}''.
\end{itemize}
\end{tcolorbox}

\section{Challenges from the Emerging Era}

Beyond the foundational issues of scalability and temporal representation, the rapid evolution of emerging technologies presents a new set of challenges.

\noindent \textbf{Healthcare AI models}
The deployment of complex, often opaque AI models in clinical settings is accelerating~\cite{qiu2024llm}.
These models may recommend treatments or predict risks, but they rarely explain their reasoning in a causally meaningful way for general users~\cite{cascella2023evaluating}.
One of the key challenges for VA is to serve as a \textit{causal auditor and explainer for AI systems' healthcare scenarios}.

\noindent \textbf{Multi-modal data}
The complexity of healthcare data is no longer just about the number of variables; it's about the variety of data types~\cite{gotz2016data}.
A patient's story is told through structured EHR events, continuous time-series from ICU monitors, high-resolution pathology images, static but massive genomic profiles, and narrative clinical notes.
A pressing new challenge is to develop VA frameworks that \textit{support causal inference across these fundamentally different modalities}.

\noindent \textbf{Collaborative inference}
Historically, VA has focused on helping users analyze data under the assumption that either no reliable causal model exists~\cite{kaul2021improving}, or the causal model is entirely pre-designed by experts~\cite{guo2023causalvis}.
A forward-looking challenge is to flip this paradigm: can we use VA to connect general users and experts to collaboratively build and refine causal models?
In such a system, the large models could analyze the data to suggest potential causal links based on statistical evidence.
The human users would then use their own knowledge or simply priors to visually confirm, reject, prune, or modify these links, adding directionality and context that the data alone cannot provide---resulting in a \textit{direct manipulation of causal models}.

\begin{tcolorbox}[
    colback=mybg, 
    colframe=myframe, 
    title={\textbf{Future Direction}},
    fonttitle=\bfseries,
    left=1mm, right=1mm, top=1mm, bottom=1mm,
    boxsep=1mm
]
\begin{itemize}[left=0pt]
    \item \textbf{Causal Auditing and Explainability for Clinical AI.} 
    A critical role for VA is to serve as a causal auditor for clinical AI. Instead of just explaining what features an AI used, VAHC research can create interfaces for ``causal sanity checks'' and support the development of causally-informed machine learning models that are more robust and reliable in real-world clinical practice.
\end{itemize}
\end{tcolorbox}

\section{The Future of Causal Inference-Enabled VAHC}

\label{sec-conclusion}

We have outlined the profound need to use visualization methods to move beyond simple correlations and toward a causality-aware understanding of health data.
We argue that the path forward is not to build automated ``causality-finding machines'' that replace human experts, but rather that we need to achieve a more ambitious goal: augmenting human intelligence to support causal reasoning through a new generation of visual analytics tools.

\acknowledgments{
We thank the reviewers for their insightful comments.
This work was supported by the National Science Foundation under Grant No. 2211845.
}

\bibliographystyle{abbrv-doi}

\bibliography{template}

\begin{thebibliography}{10}

\bibitem{borland2021enabling}
D.~Borland, I.~Brain, K.~Fecho, E.~Pfaff, H.~Xu, J.~Champion, C.~Bizon, and D.~Gotz.
\newblock Enabling longitudinal exploratory analysis of clinical covid data.
\newblock In {\em IEEE Workshop on Visual Analytics in Healthcare (VAHC)}, pp. 19--24, 2021.

\bibitem{borland2024using}
D.~Borland, A.~Z. Wang, and D.~Gotz.
\newblock Using counterfactuals to improve causal inferences from visualizations.
\newblock {\em IEEE Computer Graphics and Applications}, 44(1):95--104, 2024. doi: {{%
10\hspace{.1pt}\discretionary{.}{%
}{.}\hspace{.4pt}1109\discretionary{/}{%
}{/}MCG\hspace{.1pt}\discretionary{.}{%
}{.}\hspace{.4pt}2023\hspace{.1pt}\discretionary{.}{%
}{.}\hspace{.4pt}3338788}}


\bibitem{cascella2023evaluating}
M.~Cascella, J.~Montomoli, V.~Bellini, and E.~Bignami.
\newblock Evaluating the feasibility of chatgpt in healthcare: an analysis of multiple clinical and research scenarios.
\newblock {\em Journal of medical systems}, 47(1):33, 2023.

\bibitem{charig_comparison_1986}
C.~R. Charig, D.~R. Webb, S.~R. Payne, and J.~E. Wickham.
\newblock Comparison of treatment of renal calculi by open surgery, percutaneous nephrolithotomy, and extracorporeal shockwave lithotripsy.
\newblock {\em Brit. Med. J.}, 1986.

\bibitem{dash2019big}
S.~Dash, S.~K. Shakyawar, M.~Sharma, and S.~Kaushik.
\newblock Big data in healthcare: management, analysis and future prospects.
\newblock {\em Journal of big data}, 6(1):1--25, 2019.

\bibitem{fan2024visual}
M.~Fan, J.~Yu, D.~Weiskopf, N.~Cao, H.-Y. Wang, and L.~Zhou.
\newblock Visual analysis of multi-outcome causal graphs.
\newblock {\em IEEE Transactions on Visualization and Computer Graphics}, 31(1):656 -- 666, 2025.

\bibitem{gimenez2008complexity}
O.~Gim{\'e}nez and A.~Jonsson.
\newblock The complexity of planning problems with simple causal graphs.
\newblock {\em Journal of Artificial Intelligence Research}, 31:319--351, 2008.

\bibitem{gotz2016data}
D.~Gotz and D.~Borland.
\newblock Data-driven healthcare: challenges and opportunities for interactive visualization.
\newblock {\em IEEE Computer Graphics and Applications}, 36(3):90--96, 2016.

\bibitem{guo2023causalvis}
G.~Guo, E.~Karavani, A.~Endert, and B.~C. Kwon.
\newblock Causalvis: Visualizations for causal inference.
\newblock In {\em Proceedings of the 2023 CHI Conference on Human Factors in Computing Systems}, pp. 1--20. ACM, New York, 2023. doi: {{%
10\hspace{.1pt}\discretionary{.}{%
}{.}\hspace{.4pt}1145\discretionary{/}{%
}{/}3544548\hspace{.1pt}\discretionary{.}{%
}{.}\hspace{.4pt}3581236}}


\bibitem{haukoos2015propensity}
J.~S. Haukoos and R.~J. Lewis.
\newblock The propensity score.
\newblock {\em Jama}, 314(15):1637--1638, 2015.

\bibitem{hullman2021designing}
J.~Hullman and A.~Gelman.
\newblock Designing for interactive exploratory data analysis requires theories of graphical inference.
\newblock {\em Harvard Data Science Review}, 3(3):10--1162, 2021. doi: {{%
10\hspace{.1pt}\discretionary{.}{%
}{.}\hspace{.4pt}1162\discretionary{/}{%
}{/}99608F92\hspace{.1pt}\discretionary{.}{%
}{.}\hspace{.4pt}3AB8A587}}


\bibitem{jin2020visual}
Z.~Jin, S.~Guo, N.~Chen, D.~Weiskopf, D.~Gotz, and N.~Cao.
\newblock Visual causality analysis of event sequence data.
\newblock {\em IEEE Transactions on Visualization and Computer Graphics}, 27(2):1343--1352, 2020. doi: {{%
10\hspace{.1pt}\discretionary{.}{%
}{.}\hspace{.4pt}1109\discretionary{/}{%
}{/}TVCG\hspace{.1pt}\discretionary{.}{%
}{.}\hspace{.4pt}2020\hspace{.1pt}\discretionary{.}{%
}{.}\hspace{.4pt}3030465}}


\bibitem{julious1994confounding}
S.~A. Julious and M.~A. Mullee.
\newblock Confounding and {S}impson's paradox.
\newblock {\em Bmj}, 309(6967):1480--1481, 1994.

\bibitem{kaul2021improving}
S.~Kaul, D.~Borland, N.~Cao, and D.~Gotz.
\newblock Improving visualization interpretation using counterfactuals.
\newblock {\em IEEE Transactions on Visualization and Computer Graphics}, 28(1):998--1008, 2021. doi: {{%
10\hspace{.1pt}\discretionary{.}{%
}{.}\hspace{.4pt}1109\discretionary{/}{%
}{/}TVCG\hspace{.1pt}\discretionary{.}{%
}{.}\hspace{.4pt}2021\hspace{.1pt}\discretionary{.}{%
}{.}\hspace{.4pt}3114779}}


\bibitem{pearl2009causality}
J.~Pearl.
\newblock {\em Causality}.
\newblock Cambridge University Press, Cambridge, 2009. doi: {{%
10\hspace{.1pt}\discretionary{.}{%
}{.}\hspace{.4pt}1017\discretionary{/}{%
}{/}CBO9780511803161}}


\bibitem{qiu2024llm}
J.~Qiu, K.~Lam, G.~Li, A.~Acharya, T.~Y. Wong, A.~Darzi, W.~Yuan, and E.~J. Topol.
\newblock Llm-based agentic systems in medicine and healthcare.
\newblock {\em Nature Machine Intelligence}, 6(12):1418--1420, 2024.

\bibitem{wang2024empirical}
A.~Z. Wang, D.~Borland, and D.~Gotz.
\newblock An empirical study of counterfactual visualization to support visual causal inference.
\newblock {\em Information Visualization}, 23(2):197--214, 2024. doi: {{%
10\hspace{.1pt}\discretionary{.}{%
}{.}\hspace{.4pt}1177\discretionary{/}{%
}{/}14738716241229437}}


\bibitem{wang2024framework}
A.~Z. Wang, D.~Borland, and D.~Gotz.
\newblock A framework to improve causal inferences from visualizations using counterfactual operators.
\newblock {\em Information Visualization}, 2024. doi: {{%
10\hspace{.1pt}\discretionary{.}{%
}{.}\hspace{.4pt}1177\discretionary{/}{%
}{/}14738716241265120}}


\bibitem{wang2024beyond}
A.~Z. Wang, D.~Borland, and D.~Gotz.
\newblock Beyond correlation: Incorporating counterfactual guidance to better support exploratory visual analysis.
\newblock {\em IEEE Transactions on Visualization and Computer Graphics}, 31(1):776 -- 786, 2025.

\bibitem{wang2015visual}
J.~Wang and K.~Mueller.
\newblock The visual causality analyst: An interactive interface for causal reasoning.
\newblock {\em IEEE Transactions on Visualization and Computer Graphics}, 22(1):230--239, 2015. doi: {{%
10\hspace{.1pt}\discretionary{.}{%
}{.}\hspace{.4pt}1109\discretionary{/}{%
}{/}TVCG\hspace{.1pt}\discretionary{.}{%
}{.}\hspace{.4pt}2015\hspace{.1pt}\discretionary{.}{%
}{.}\hspace{.4pt}2467931}}


\bibitem{wang2017visual}
J.~Wang and K.~Mueller.
\newblock Visual causality analysis made practical.
\newblock In {\em 2017 IEEE Conference on visual analytics science and technology (VAST)}, pp. 151--161. IEEE, 2017.

\end{thebibliography}
\end{document}